\documentclass[11pt,a4paper]{article}
\usepackage{amssymb} \usepackage{amsmath} \usepackage{graphicx}
\usepackage{epsfig,latexsym}
\begin{document}

\def\bef{\begin{figure}}
\def\eef{\end{figure}}
\newcommand{\ans}{ansatz }
\newcommand{\be}[1]{\begin{equation}\label{#1}}
\newcommand{\beq}{\begin{equation}}
\newcommand{\ee}{\end{equation}}
\newcommand{\beqn}[1]{\begin{eqnarray}\label{#1}}
\newcommand{\eeqn}{\end{eqnarray}}
\newcommand{\bd}{\begin{displaymath}}
\newcommand{\ed}{\end{displaymath}}
\newcommand{\mat}[4]{\left(\begin{array}{cc}{#1}&{#2}\\{#3}&{#4}
\end{array}\right)}
\newcommand{\matr}[9]{\left(\begin{array}{ccc}{#1}&{#2}&{#3}\\
{#4}&{#5}&{#6}\\{#7}&{#8}&{#9}\end{array}\right)}
\newcommand{\matrr}[6]{\left(\begin{array}{cc}{#1}&{#2}\\
{#3}&{#4}\\{#5}&{#6}\end{array}\right)}
\newcommand{\cvb}[3]{#1^{#2}_{#3}}
\def\lsim{\raise0.3ex\hbox{$\;<$\kern-0.75em\raise-1.1ex
e\hbox{$\sim\;$}}}
\def\gsim{\raise0.3ex\hbox{$\;>$\kern-0.75em\raise-1.1ex
\hbox{$\sim\;$}}}
\def\abs#1{\left| #1\right|}
\def\simlt{\mathrel{\lower2.5pt\vbox{\lineskip=0pt\baselineskip=0pt
           \hbox{$<$}\hbox{$\sim$}}}}
\def\simgt{\mathrel{\lower2.5pt\vbox{\lineskip=0pt\baselineskip=0pt
           \hbox{$>$}\hbox{$\sim$}}}}
\def\unity{{\hbox{1\kern-.8mm l}}}
\newcommand{\eps}{\varepsilon}
\def\ep{\epsilon}
\def\ga{\gamma}
\def\Ga{\Gamma}
\def\om{\omega}
\def\omp{{\omega^\prime}}
\def\Om{\Omega}
\def\la{\lambda}
\def\La{\Lambda}
\def\al{\alpha}
\newcommand{\ov}{\overline}
\renewcommand{\to}{\rightarrow}
\renewcommand{\vec}[1]{\mathbf{#1}}
\newcommand{\vect}[1]{\mbox{\boldmath$#1$}}
\def\tm{{\widetilde{m}}}
\def\mcirc{{\stackrel{o}{m}}}
\newcommand{\Dm}{\Delta m}
\newcommand{\dm}{\varepsilon}
\newcommand{\tanb}{\tan\beta}
\newcommand{\nbar}{\tilde{n}}
\newcommand\PM[1]{\begin{pmatrix}#1\end{pmatrix}}
\newcommand{\up}{\uparrow}
\newcommand{\down}{\downarrow}
\def\omE{\omega_{\rm Ter}}
%
%%%%%%%%%%     mauri    %%%%%%%%%%%%%%%%%%%%%%%%%%%%%%%%%

\newcommand{\Dsusy}{{susy \hspace{-9.4pt} \slash}\;}
\newcommand{\DCP}{{CP \hspace{-7.4pt} \slash}\;}
\newcommand{\mc}{\mathcal}
\newcommand{\gr}{\mathbf}
\renewcommand{\to}{\rightarrow}
\newcommand{\gtc}{\mathfrak}
\newcommand{\wh}{\widehat}
\newcommand{\br}{\langle}
\newcommand{\kt}{\rangle}

%%%%%%%%%%%%%%%%%%%%%%%%%%%%%%%%%%%%%%%%%%%%%%%%%%%%%%%%%%

% barbara Ricci  %definizione di minore e maggiore simile
\def\lsim{\mathrel{\mathop  {\hbox{\lower0.5ex\hbox{$\sim$}
\kern-0.8em\lower-0.7ex\hbox{$<$}}}}}
\def\gsim{\mathrel{\mathop  {\hbox{\lower0.5ex\hbox{$\sim$}
\kern-0.8em\lower-0.7ex\hbox{$>$}}}}}
%%%%%%%%%%%%%%%%%%%%%%%%%%%%%%%%%%

\def\nn{\\  \nonumber}
\def\de{\partial}
\def\brf{{\mathbf f}}
\def\bbf{\bar{\bf f}}
\def\bF{{\bf F}}
\def\bbF{\bar{\bf F}}
\def\bA{{\mathbf A}}
\def\bB{{\mathbf B}}
\def\bG{{\mathbf G}}
\def\bI{{\mathbf I}}
\def\bM{{\mathbf M}}
\def\bY{{\mathbf Y}}
\def\bX{{\mathbf X}}
\def\bS{{\mathbf S}}
\def\bb{{\mathbf b}}
\def\bh{{\mathbf h}}
\def\bg{{\mathbf g}}
\def\bla{{\mathbf \la}}
\def\bmu{\mathbf m }
\def\by{{\mathbf y}}
\def\bmu{\mbox{\boldmath $\mu$} }
\def\bsig{\mbox{\boldmath $\sigma$} }
\def\bunity{{\mathbf 1}}
\def\cA{{\cal A}}
\def\cB{{\cal B}}
\def\cC{{\cal C}}
\def\cD{{\cal D}}
\def\cF{{\cal F}}
\def\cG{{\cal G}}
\def\cH{{\cal H}}
\def\cI{{\cal I}}
\def\cL{{\cal L}}
\def\cN{{\cal N}}
\def\cM{{\cal M}}
\def\cO{{\cal O}}
\def\cR{{\cal R}}
\def\cS{{\cal S}}
\def\cT{{\cal T}}
\def\eV{{\rm eV}}
%
%%%%%%%%%%%%%%%%%%%%%%%%%%%%%%%%%%%%%

%%%%%%%%%%%%%% titlepage
%%%%%%%%%%%%%% titlepage
%%%%%%%%%%%%%% titlepage
%%%%%%%%%%%%%% titlepage

%\begin{flushright}
%DFAQ-2001/TH/02\\
%arXiv: hep-ph/0507031 \\
%March 2001
%\today

%\end{flushright}
%\vspace{1.0cm}

\large
 \begin{center}
 {\Large \bf Unitarization and Causalization of Non-local quantum field theories by Classicalization}
%\vspace{3mm}
%Ultra High Energy Cosmic Rays }
%Phenomenological and Astrophysical implications }
 \end{center}

 \vspace{0.1cm}

 \vspace{0.1cm}
 \begin{center}
{\large Andrea Addazi}\footnote{E-mail: \,  andrea.addazi@infn.lngs.it} \\
% \begin{center}
 %\vskip 0.1 truecm
 %\centerline
 %{$^a$
{\it \it Dipartimento di Fisica,
 Universit\`a di L'Aquila, 67010 Coppito, AQ \\
LNGS, Laboratori Nazionali del Gran Sasso, 67010 Assergi AQ, Italy}
\end{center}

\vspace{1cm}
\begin{abstract}
\large

We suggest that classicalization can cure non-local quantum field theories
from acausal divergences in scattering amplitudes, restoring unitarity and causality.
In particular, in "trans-non-local" limit, the formation of non-perturbative classical configurations, called {\it classicalons},
in scatterings like $\phi\phi\rightarrow \phi\phi$,
can avoid typical acausal divergences.
\end{abstract}

\baselineskip = 20pt

\section{Introduction}

Classicalization is an alternative "road" to an UV completion
 of quantum field theories with respect to Wilsonian one. Such a phenomena 
 was conjectured by Dvali and collaborators, 
 and a lot of examples seem to sustain this 
 argument \cite{Dvali1}. 
 In this paper, we would like to argument possible connections 
among classicalization and non-local quantum field theories. 
To formulate a consistent quantum field theory 
without the locality principle is an old problem:
it is well known as an "insidious problem". 
 In fact, even if one can formulate a consistent model at tree level, 
 unitarity and causality will be inevitably lost  at quantum level \cite{W1,W2,Moffat,W3,Evens,Joglekar1,Joglekar2}
. For example, in a non-local scalar field theory, 
 acausalities will inevitably appear in scatterings like $\phi\phi \rightarrow \phi\phi$
 or $\phi\phi \rightarrow \phi\phi\phi\phi$.
 However, classicalization can offer a natural way-out 
 to acausal divergences: the formation of classical extended 
 objects of radius $R>\Lambda_{NL}^{-1}$ in scatterings with $E_{CM}>\Lambda_{NL}$, named classicalons, 
 naturally avoids these infinities (where $\Lambda_{NL}$ is the Non-locality effective scale). 
 In particular, we will quantitatively focus on a particularly promising class of Non-local models 
studied by Eliezer, Woodard, Moffat, Kleppe, Evens and Joglekar \cite{W1,W2,Moffat,W3,Evens,Joglekar1,Joglekar2}
\footnote{An alternative approach to non-local QFT and non-local quantum gravity is considered in 
\cite{Briscese:2012ys,Modesto:2013ioa,Bambi:2013caa,Modesto:2013oma,Bambi:2014uua,Modesto:2014xta,Modesto:2014lga,Calcagni:2014vxa,Modesto:2015lna,Dona:2015tra}. The main difference 
with respect to our approach is the following: a infinite number of local gauge transformations
are considered in their case rather than a non-linear one. I am grateful to Leonardo Modesto for discussions
on these aspects. 
}.

\section{EWMKEJ's model: nonlocal scalar field theory}

Let us briefly review the 
 EWMKEJ ({\it i.e.}, Eliezer, Woodard, Moffat, Kleppe, Evens, Joglekar)   
model of a nonlocal $\lambda \phi^{4}$ theory. 
We start from an action
\begin{equation}
\label{W1}
S[\phi]=\mathcal{F}[\phi]+\mathcal{I}[\phi],
\end{equation}
where $\mathcal{F}[\phi]$ is the free part, $\mathcal{I}[\phi]$ is the interaction part.
We suppose analytic functional around the vacuum.
$\mathcal{F}$ has a general form 
\begin{equation}
\label{W2}
\mathcal{F}[\phi]=-\frac{1}{2}\int {\rm d}^{4}x\phi_{i}F_{ij}\phi_{j}.
\end{equation}
The action $S$ is "nonlocalized"  through a "smearing operator" $\mathcal{E}$.
The EWMKEJ choice corresponds to an exponential smearing operator: 
\begin{equation}
\label{W3}
\mathcal{E}=\rm exp\left[\frac{F}{2\Lambda_{NL}^{2}}\right],
\end{equation}
$\Lambda_{NL}$ is an effective scale of non-locality. 
 Then, $\phi$ are smeared as
\begin{equation}
\label{W2}
\hat{\phi}_{i}=\mathcal{E}_{ij}^{-1}\phi_{j}.
\end{equation}
Let us define the operator
\begin{equation}
\label{W3}
\mathcal{K} \equiv (\mathcal{E}^{2}-I)F^{-1}.
\end{equation}
Now, we introduce an auxiliary field $\varphi_{i}$ for each matter field $\phi_{i}$:
\begin{equation}
\label{W4}
\hat{S}[\phi,\varphi]=\mathcal{F}[\hat{\phi}]-\mathcal{A}[\varphi]+\mathcal{I}[\phi+\varphi],
\end{equation}
\begin{equation}
\label{W5}
\mathcal{A}[\varphi]=-\frac{1}{2}\int {\rm d}^{4}x\varphi_{i} \mathcal{K}_{ij}^{-1}\varphi_{j}.
\end{equation}
The classical auxiliar field equation is
\begin{equation}
\label{W6}
\frac{\delta \hat{S}[\phi, \varphi]}{\delta \varphi(x)}=0.
\end{equation}
The final 
nonlocal action is non-linearly obtained 
by  substituting the solution of  Eq. (\ref{W6}), into Eq. (\ref{W4});
{\it i.e} by substituting 
$\varphi_{i}=\mathcal{K}_{ij} {\delta \mathcal{I}[\phi+\varphi] \over \delta \varphi_{j}}$.

\subsection{Gauge symmetries}

How can be constructed a local gauge symmetry in a non-local model?
Gauge symmetry can be encoded in 
a nonlocal theory with a new nonlinear transformation 
rule. In fact, as shown in Ref.\cite{W3} for the scalar theory, 
if an infinitesimal transformation
$\delta \phi_{i}=T_{i}[\phi]$
generates a symmetry of the local action $S[\phi]$, then a transformation
$\hat{\delta}\phi_{i}=\mathcal{E}_{ij}^{2}T_{j}[\phi+\varphi[\phi]]$
generates a symmetry for the corresponding nonlocal action $\hat{S}[\phi]$. 
In a broad sense, the procedure for obtaining a nonlocal theory preserves 
a deformed version of the usual continuous symmetry, and we can write 
\begin{equation} 
\label{auxialiar}
\hat{\delta}\varphi_{i}[\phi]=\left(I-\mathcal{E}^{2}\right)_{ij}T_{j}\left[\phi
+\varphi[\phi]\right]-K_{ij}\left[\phi+\varphi[\phi]\right]\frac{\delta T_{k}}
{\delta \phi_{j}}\left[\phi+\varphi[\phi] \right]\mathcal{E}_{kl}^{2}
\frac{\delta \hat{S}[\phi]}{\delta \phi_{l}},
\end{equation}
\begin{equation} 
\label{Ktr}
K_{ij}^{-1}[\phi]=\mathcal{K}_{ij}^{-1}-\frac{\delta^{2}\mathcal{I}[\phi]}{\delta \phi_{i}\delta \phi_{j}}.
\end{equation}

\subsection{Quantization in the EWMKEJ model}

Consider  the vacuum expectation value of an arbitrary operator 
$\mathcal{O}$:
\begin{equation}
\label{quantization}
\langle \mathcal{T}\left(\mathcal{O}[\phi]\right) \rangle_{\mathcal{E}}=\int \mathcal{D}\phi\, m[\phi]
\left(GF\right)\,\mathcal{O}[\hat{\phi}]\,\rm e^{i\hat{S}[\phi]}
\end{equation}
($\mathcal{T}$ is the time-ordering operator, and $GF$ is the gauge fixing). 
In this definition, 
$\mathcal{O}$ is nonlocally regulated and
eq.(\ref{quantization}) defines the quantization in non-local model.

A consistent quantization of EWMKEJ 
requests the  existence of 
the measure factor $m[\phi]$ and the gauge fixing.
A measure functional is necessary 
 in order to preserve unitarity. 
Unitarity of a nonlocal quantum field theory was discussed 
 in papers cited above.
 In particular,
a large subspace $\mathcal{M}$ in the Fock space
posses unitarity at three level. As in local QFT,
$\mathcal{M}$ can also have unphysical polarizations
as BRST ghost fields. On the other hand, non-linear
gauge invariance guarantees BRST ghosts' decoupling  
on shells.

So, the EWMKEJ procedure starts 
from a local QFT 
in order to obtain a non-local deformation. 
Generically, the starting QFT has 
continuos  transformations
$\delta \phi_{i}=T_{i}[\phi]$ of the local action $S[\phi]$.
EWMKEJ procedure generates 
 corresponding transformations 
$\hat{\delta}\phi_{i}=\mathcal{E}_{ij}^{2}T_{j}[\phi+\varphi[\phi]]$ for the non-local QFT. 
However, this transformation has to 
preserve $\mathcal{D}\phi \, m[\phi]$, 
{\it i.e} $\hat{\delta} [\mathcal{D}\phi \, m[\phi]]=0$. 
Such a condition corresponds to
\begin{equation} \label{conq}
\hat{\delta}\left[ {\rm log}(m[\phi]) \right]=-{\rm Tr}\left[\frac{\delta \hat{\delta} \phi_{i}}
{\delta \phi_{j}}\right]=-{\rm Tr} \left[\mathcal{E}_{ik}^{2}\frac{\delta T_{k}}
{\delta \phi_{l}}[\phi+\varphi[\phi]]K_{lk}[\phi+\varphi[\phi]]\mathcal{K}_{kj}^{-1} \right].
\end{equation}

Under this quantization procedure, we can recover 
Feynman rules of $\hat{S}[\phi,\varphi]$ as simple extension of usual ones:
propagators are smeared by a factor $\mathcal{E}^{2}$, 
as mentioned above. The $\varphi$ are auxiliary fields 
propagating only off-shell, because they are projected-out 
by solutions of classical field equations $\varphi[\phi]$. 
 
\subsection{Non-local Feynman rules}

The "funny trick" of the auxiliar field 
allows to obtain simple 
Feynman rules, as understood deformations 
of usual ones. 
Let us resume the new prescriptions demonstarted in papers cited above: 

i) the vertices are unchanged;

ii) the smeared propagator for the fields $\phi$ reads as 
\begin{equation} 
\label{pro1}
-\frac{{\rm i}\mathcal{E}^{2}}{(F+{\rm i}\epsilon)}
\end{equation}

iii) The smeared propagators for the auxiliary\
fields $\varphi$ are
\begin{equation} 
\label{pro2}
-\frac{{\rm i}[I-\mathcal{E}^{2}]}{(F+{\rm i}\epsilon)}
\end{equation}
where $I$ is the identity-operator. 

Let us choice 
\begin{equation} 
\label{b1}
F=\Box+m^{2}
\end{equation}

We can conveniently write 
Feynman rules in momentum space are as follows.

ii)
\begin{equation} 
\label{prop1psi1}
{\rm i}\frac{ {\rm exp}\left( \frac{-(p^{2}-m^{2})}{\Lambda_{NL}^{2}}\right) } {(p^{2}-m^{2}+{\rm i}\epsilon)}
\end{equation}

iii)
\begin{equation} 
\label{prop2psi2}
{\rm i}\frac{ \left[I-{\rm exp}\left( \frac{-(p^{2}-m^{2})}{\Lambda_{NL}^{2}}\right) \right]}
{(p^{2}-m^{2}+{\rm i}\epsilon)}
\end{equation}

\subsection{Bogoliubov-Shirkov causality conditions}
In order to test causality and unitarity at all the orders
of the pertubation series,
one can find recursive relations on the S-matrices.
Such conditions were discussed by Bogoliubov and Shirkov 
in their book on QFT \cite{BS}. 

\subsubsection{Unitarity and Causality: definitions}

It is useful to remind what unitarity and causality 
impose on the S-matrix. 

 Unitarity means that the total probability of processes
equal to one: $S$-matrix has to satisfy the condition 
\begin{equation}
\label{Smatrix}
{\mathcal S}{\mathcal S}^{\dagger}=
{\mathcal S}^{\dagger}{\mathcal S}=I.
\end{equation}

Causality can be reformulated as the cluster decomposition principle
on the S-matrix \cite{Weinberg}.
If multi-particle transitions $A_{1}\rightarrow B_{1},\,\,A_{2}\rightarrow B_{2},...,\,\,A_{n}\rightarrow B_{n}$ 
are studied in $N$ different laboratories, with different positions $z_{1,..,n}$ in the space-time with
$(z_{i}-z_{j})^{2}<0$ ($i\neq j$, $i,j=1,...,n$), 
then the S-matrix will be decomposed as
\begin{equation} 
\label{Cluster}
\mathcal{S}_{B_{1}+B_{2}+...B_{n},A_{1}+A_{2}+...+A_{n}}
=\mathcal{S}_{B_{1}A_{1}}\mathcal{S}_{B_{2}A_{2}}...\mathcal{S}_{B_{n}A_{n}}.
\end{equation}
Rel.(\ref{Cluster}), is strictly connected to 
the hypothesis that 
quantum fields 
commute as
\begin{equation} 
\label{phiphi}
[\phi(z_{i}),\phi(z_{j})]=0,\,\,\,\,\,(z_{i}-z_{j})^{2}<0,
\end{equation}
which is equivalent to a microcausality condition on the S-matrix:
\begin{equation} 
\label{transl}
\frac{\delta}{\delta \phi(z_{i})}\left(\frac{\delta {\mathcal S}[\phi]}{\delta \phi(z_{j})}
{\mathcal S}^{\dagger}[\phi] \right)=0,\,\,\,\,for\,\,z_{i}< z_{j}.
\end{equation}

\subsubsection{Recursive relations as a "Test-Bed" for unitarity and causality}

Let us perform a Dyson expansion of the S-matrix
with respect to the couplings $c(z)$ promoted to spacetime fields \cite{BS}:
\begin{equation} 
\label{Sexp}
\mathcal{S}[c(x)]=I+\sum_{n=1}^{\infty}\frac{1}{n!}\int 
{\rm d}z_{1}...{\rm d}z_{n}\mathcal{T}\{{\cal S}_{n}(z_{1},...,z_{n})c(z_{1})...c(z_{n})\}.
\end{equation}

Let us note that
conditions (\ref{Smatrix}) and (\ref{transl}) can be
rewritten in terms of $c(z)$, 
through a functional Legendre transform $\phi(z)\rightarrow c(z)$.  
Then, for $c(z)\rightarrow c=cost$, we can insert the expansion (\ref{Sexp}) into (\ref{Smatrix}) and (\ref{transl}),
reverting the usual expansion. 
These allow to obtain the 
 following recursive relations, for perturbation theory:
\begin{equation} 
\label{norder}
{\mathcal R}_{n}={\rm i}{\cal S}_{n+1}(y,z_{1},...,z_{n})+{\rm i}\sum_{0\leq k \leq n-1}
\mathcal{P}\{{\cal S}_{k+1}(y,z_{1},..,z_{k}){\cal S}_{n-k}^{\dagger}(z_{k+1},...,z_{n})\} ,
\end{equation}
and 
\begin{equation} 
\label{norder2}
{\cal S}_{n}(z_{1},...,z_{n})+ {\cal S}^{\dagger}_{n}(z_{1},...,z_{n})
+\sum_{1\leq k \leq n-1}
{\mathcal P}\{{\cal S}_{k}(z_{1},..,z_{k}){\mathcal S}_{n-k}^{\dagger}(z_{k+1},...,z_{n})\}=0,
\end{equation}
where 
 $\mathcal{P}\{ \}$ is the sum over all partitions of $\{z_{1},...,z_{n}\}$ into
$k$ and $n-k$ elements. 
For example, $\{z_{1},..,z_{k}\}$, $\{z_{k+1},..,z_{n}\}$ and so on. 

The causality conditions for the first two orders of the perturbation theory are
\begin{equation} 
\label{cc}
{\mathcal R}_{1}(x,y)={\rm i}\Bigr[{\cal S}_{2}(x,y)+{\cal S}_{1}(x){\cal S}_{1}^{\dagger}(y)\Bigr]=0,
\end{equation}
\begin{equation} 
\label{cc2}
{\mathcal R}_{2}(x,y)={\rm i}\Bigr[{\cal S}_{3}(x,y,z)+{\cal S}_{1}(x){\cal S}_{2}^{\dagger}(y,z)
+{\cal S}_{2}(x,y){\cal S}_{1}^{\dagger}(z)+{\cal S}_{2}(x,z){\cal S}_{1}^{\dagger}(y)\Bigr]=0.
\end{equation}
On the other hand, unitarity is expressed at the first order as
\begin{equation} 
\label{un1}
{\mathcal S}_{1}(x)+{\mathcal S}_{1}^{\dagger}(x)=0,
\end{equation}
\begin{equation} 
\label{un2}
{\mathcal S}_{2}(x,y)+{\mathcal S}_{2}^{\dagger}(x,y)
+{\mathcal S}_{1}(x){\mathcal S}_{1}^{\dagger}(y)+{\mathcal S}_{1}(y){\mathcal S}_{1}^{\dagger}(x)=0.
\end{equation}
An alternative useful way to rewrite (\ref{cc})-(\ref{cc2}) is
\begin{equation} 
\label{C1}
\mathcal{R}_{1}=\int {\rm d}^{4}x {\rm d}^{4}y
[\theta(x_{0}-y_{0})\mathcal{R}_{1}(x,y)+\theta(y_{0}-x_{0})\mathcal{R}_{1}(y,x)]=0,
\end{equation}
\begin{equation} 
\label{C2}
\mathcal{R}_{2}=\int {\rm d}^{4}x {\rm d}^{4}y {\rm d}^{4}z 
\mathcal{R}_{2}(x,y,z)\theta(x_{0}-y_{0})\theta(y_{0}-z_{0})+ 5 \; {\rm symmetric}\, 
{\rm terms} =0.
\end{equation}

Let us remark that these bounds are surely a "test-bed" for causality and unitarity violations. 
In fact, suppose to calculate a 1 and 2 vertices' amplitudes:
from the 
 {\it momentum dependence of $\mathcal{S}_{1,2}$}, implying a certain function of momentum 
in $\mathcal{R}_{1}$, {\it one can manifestly see signals of causality or unitarity violations}. 
In fact, if the net contribution of the relevant amplitudes to $\mathcal{R}_{1}$ is proportional to a polynomial function of the Mandelstam 
variables $\mathcal{X}^{n}/\Lambda_{NL}^{2n}$ with $n>1$ $(\mathcal{X}=s,t,u)$, for $\mathcal{X}>>\Lambda_{NL}^{2}$
a breakdown of causality will occur. 
In order to have a causal theory, $\mathcal{R}_{1}$ has to be zero, as well as $\mathcal{R}_{2,...,n}$. 
Analogous for unitarity. 
This is last condition is violated in 
non-local models like EWMKEJ, as we will see in the next sections. 
In fact, the presence of off-shell auxiliar fields will introduce extra un-balanced contributions violating 
causality and unitarity bounds.

\section{Acausal divergences in Scatterings and Classicalization}

\subsection{One-loop acausal diagrams in $\phi\phi \rightarrow \phi\phi$ scatterings}

\begin{figure}[t]
\centerline{ \includegraphics [height=3cm,width=0.3 \columnwidth]{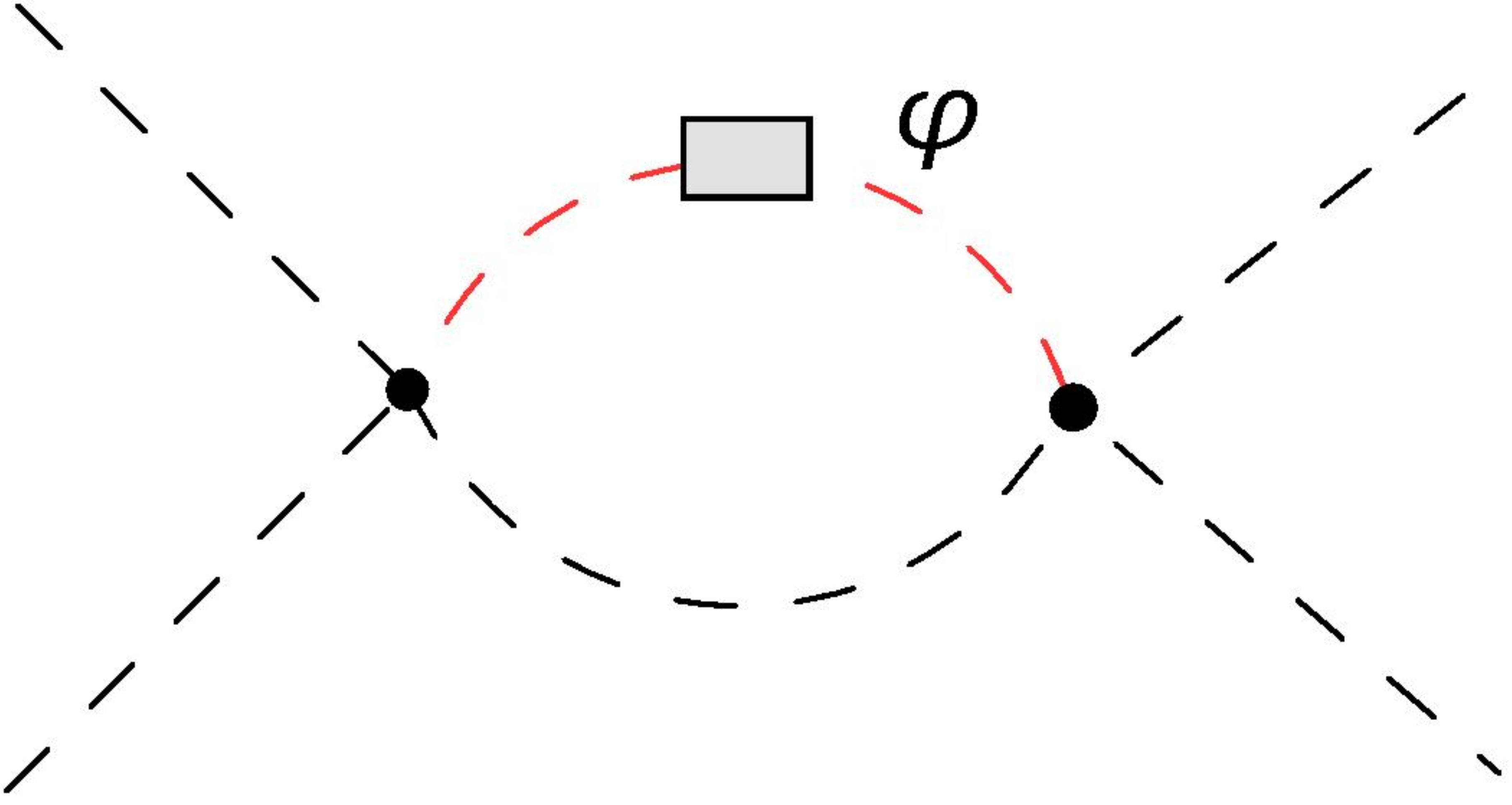}}
\vspace*{-1ex}
\caption{Example of 1-loop contribution to $\phi\phi \rightarrow \phi\phi$ scattering leading to acausal divergences.
In this diagram, one off-shell auxial field $\varphi$ (red dashed lines with a box) and an one ordinary one are inside the one-loop. This diagram is related to the interaction vertex $\lambda\phi^{3} \varphi$ inside $\mathcal{I}[\phi+\varphi]$.}
\label{plot}   % \ref{plot}
\end{figure}

In this section, we show an explicit example of how unitarity and causality 
are lost at quantum level. In particular, we briefly review 
the case of a scattering $\phi\phi\rightarrow \phi\phi$ 
corrected by a loop of an auxiliar field $\varphi$ and one field $\phi$. 
This example was discussed in Ref.\cite{Joglekar1,Joglekar2}, and Ref.\cite{Addazi:2015dxa}
in a susy generalization. 

By assuming the massless case $m=0$, the (renormalized) amplitude is
\begin{equation} 
\label{Amplitude1}
\mathcal{A}(s,t,u)=\frac{9\lambda^{2}}{4\pi^{2}}\sum_{\mathcal{X}=s,t,u}\sum_{n=0}^{\infty}a_{n}
\left(\frac{\mathcal{X}}{\Lambda_{NL}^{2}}\right)^{n}
\end{equation}
where
$$a_{n}=\frac{1}{2^{n}n(n+1)!}\left(2^{n}-1\right)$$
and $\mathcal{X}=s,t,u$ are the Mandelstam variables.

Let us comment that the reintroduction of the mass parameter will 
complicate the form of (\ref{Amplitude1}), but essentially this is not important
for our purposes. Anyway, one can hand also masses using Schwinger 
parameters, as usually done for local QFT. 
One can rewrite the amplitudes as
the following integrals in $s$-channel: 
\begin{equation} 
\label{boson}
\mathcal{A}(s)=\frac{9\lambda^{2}}{4\pi^{2}}\int_{0}^{1/2}{\rm d}x
\int_{\frac{1}{(1-x)}}^{\frac{1}{x}}\frac{{\rm d}\zeta}{\zeta}{\rm exp}
\left\{- \frac{\zeta}{\Lambda_{NL}^{2}}(m^{2}-x(1-x)s) \right\} .
\end{equation}
The complete expansion is complicated, but
we can perform an asymptotic expansion around $s=0$,
as 
\begin{equation}
\label{Amplitudeesp}
\mathcal{A}(s) \sim \sum_{n=0}^{2}a_{n}(m,\Lambda_{NL})s^{n}+{\rm O}(s^{3}),
\end{equation}
in which the coefficients 
 $a_{0,1,2}$ are expressed as the following integrals
\begin{equation} 
\label{0th}
a_{0}(m,\Lambda_{NL})=\frac{9\lambda^{2}}{4\pi^{2}}\int_{0}^{1/2}{\rm d}x\int_{\frac{1}{(1-x)}}^{1 \over x}
\frac{{\rm d}\zeta}{\zeta}{\rm e}^{-\frac{m^{2}\zeta}{\Lambda_{NL}^{2}}},
\end{equation}
\begin{equation} 
\label{1th}
a_{1}(m,\Lambda_{NL})=\frac{9\lambda^{2}}{4\pi^{2}}\int_{0}^{1/2}{\rm d}x\int_{\frac{1}{(1-x)}}^{1 \over x}
{\rm d}\zeta \frac{x(1-x)}{\Lambda_{NL}^{2}}{\rm e}^{-\frac{m^{2}\zeta}{\Lambda_{NL}^{2}}},
\end{equation}
\begin{equation} 
\label{2th}
a_{2}(m,\Lambda_{NL})=\frac{9\lambda^{2}}{4\pi^{2}}\int_{0}^{1/2}{\rm d}x\int_{\frac{1}{(1-x)}}^{1 \over x}
\zeta {\rm d}\zeta \frac{x^{2}(1-x)^{2}}{2\Lambda_{NL}^{4}}{\rm e}^{-\frac{m^{2}\zeta}{\Lambda_{NL}^{2}}}.
\end{equation}
The zeroth and first order 
of the expansion (\ref{Amplitudeesp})
 can be cancelled in the renormalization procedure, because they are just constants. 

Let us note that new polynomial terms "strongly" violate unitarity and causality relations 
in the "trans-nonlocal regime" $E>>\Lambda_{NL}$ (or $\mathcal{X}>>\Lambda_{NL}^{2}$).
In fact, such a scattering contributes only to $\int dz_{1}dz_{2}\mathcal{S}_{2}(z_{1},z_{2})$
and not to $\int dz_{1}dz_{2}\mathcal{T}\{\mathcal{S}_{1}(z_{1})\mathcal{S}^{\dagger}(z_{2})\}$:
the net $\mathcal{R}_{1}(\mathcal{X})$ is momentum-dependent as $\sim \mathcal{X}^{2}/\Lambda_{NL}^{4}+O(\mathcal{X}^{3}/\Lambda_{NL}^{6})$. This contribution is not balanced by other correspondent ones. 
 However, this conclusion is right in the Wilsonian UV completion,
 but not for a UV Classicalization. In fact, even if $E>\Lambda_{NL}$, 
 the formation of a non-perturbative extended classic object called "classicalon"
 can cutoff the minimal length probed by the scattering 
 to the size $R$ of the classicalon. 
 If this size $R>\Lambda_{NL}^{-1}$, unitarity and causality are not lost in these channels. 
 
\subsection{Classicalization}
The qualitative idea of classicalization is the following, 
in the limits of $\sqrt{s}>\Lambda_{critical}=l_{critical}^{-1}$ (as well as limits in t- and u- channels), 
(where $\Lambda_{critical},l_{critical}$ are a critical energy and length scales respectively) 
the production
of an extended classical configurations,  called "classicalon"
starts dominating the high energy amplitudes. 
For example in a $2\rightarrow 2$ process, energy-divergences
are exponentially suppressed:
new channels (shown in Fig.2)
$$2 \rightarrow Classicalon \rightarrow N$$
start to dominate for $\sqrt{s}>\Lambda_{critical}$. 
As a consequence, in the limit of $\sqrt{s}>\Lambda_{critical}$,
the scattering cannot proceed down to scales $l<l_{critical}$, 
but it will sustain 
the creation of an extended object of size $R>l_{critical}$. 
As a consequence, a scattering with $\sqrt{s}>>\Lambda_{critical}$
cannot probe distances $l<<R$!
Clearly, this will signify that if classical production dominates in our 
non-local QFT, length-scales down the non-locality length-scale will never be 
reached, {\it i.e} no-break down of unitarity and causality in scattering amplitudes!

\begin{figure}[t]
\centerline{ \includegraphics [height=5cm,width=0.5 \columnwidth]{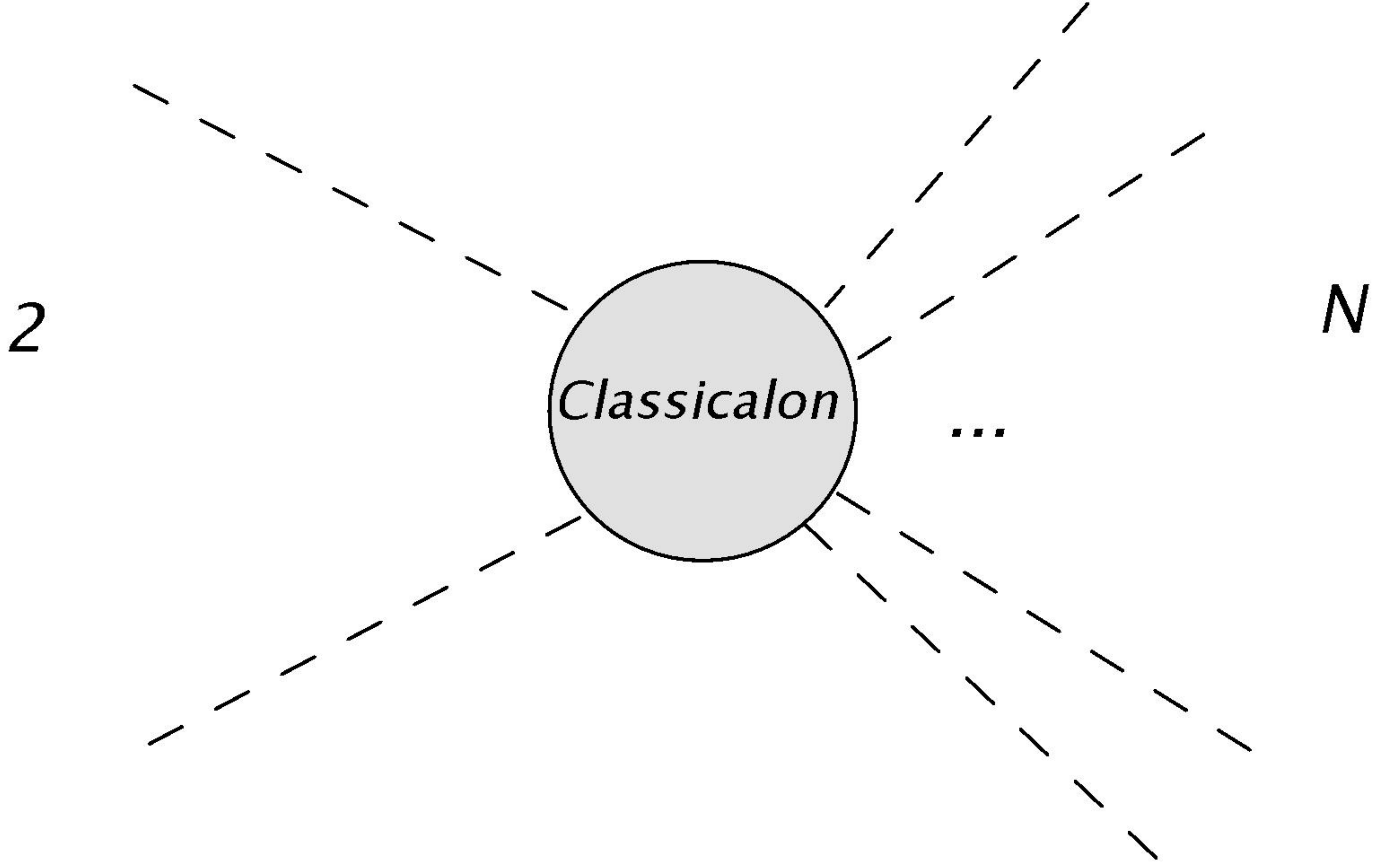}}
\vspace*{-1ex}
\caption{Classicalon' production in a scattering of two particles.}
\label{plot}   % \ref{plot}
\end{figure}

\begin{figure}[t]
\centerline{ \includegraphics [height=6cm,width=0.7 \columnwidth]{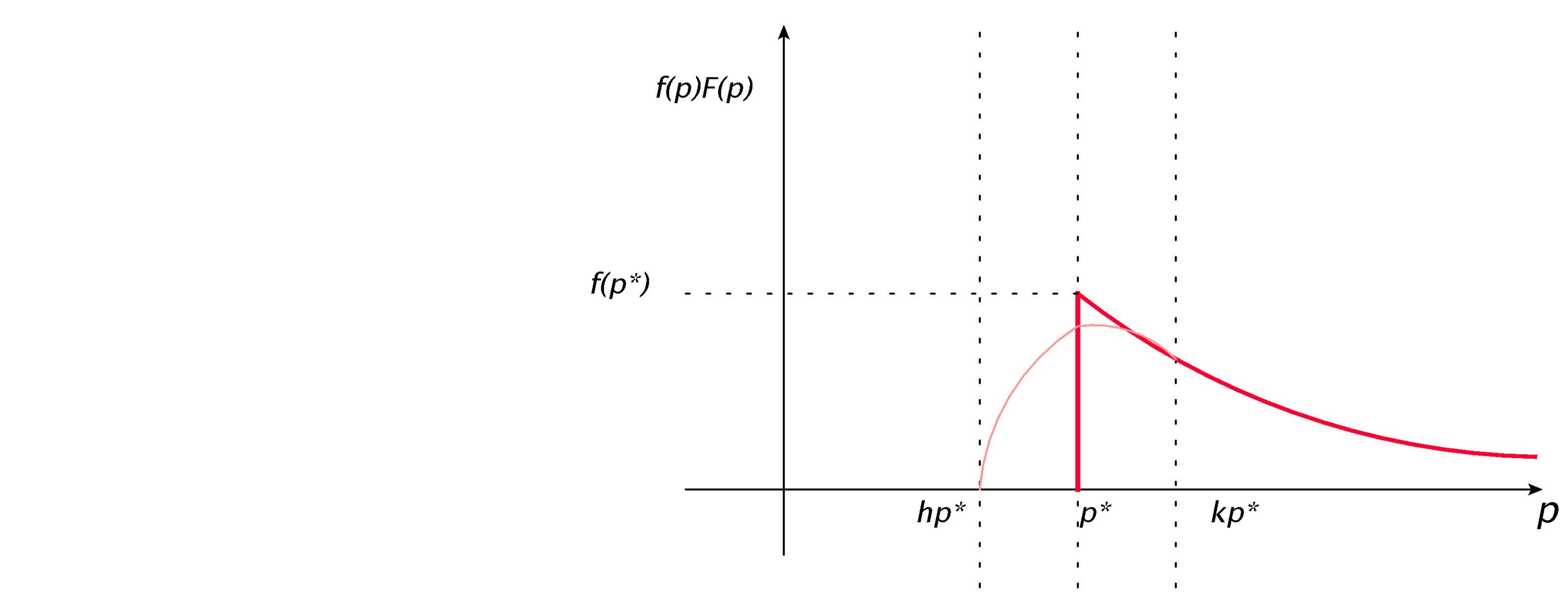}}
\vspace*{-1ex}
\caption{
Qualitative examples of $f(p)F(p)$ distributions (where $F(p)=\mathcal{F}(p-p^{*})$):
in thick red the one with ansatz $\mathcal{F}=\Theta(p-p^{*})$ 
while in light red and example of 
smoothed distributions converging to the Heaviside 
case for $p>kp^{*}$ and $p<hp^{*}$.}
\label{plot}   % \ref{plot}
\end{figure}

Since we are talking about the formation of a classical configuration, 
one of the most appropriate languages, in order to describe classicalons,
is the path integral formalism \cite{Class1}.
The idea is to study quantum fluctuations 
around classicalon rather then around the trivial vacuum\footnote{In a broad sense, 
classicalons can be considered as "brothers" of solitons, instantons and other non-perturbative solutions. 
Other intriguing implications of non-perturbative solutions called "exotic instantons" were recently studied 
in \cite{Addazi:2014ila,Addazi:2014ila,Addazi:2015ata,Addazi:2015rwa,Addazi:2015hka,Addazi:2015eca,Addazi:2015fua,Addazi:2015oba,Addazi:2015goa,Addazi:2015yna,Addazi:2015ewa}.}.
We will argument through path integral language how
classicalons can unitarize and causalize 
scattering amplitudes in non-local models. 
Let us explicitly rewrite the lagrangian as 
\begin{equation}
\label{L2}
\mathcal{L}=-\frac{1}{2}e^{-\frac{1}{2}\frac{\Box+m^{2}}{\Lambda_{NL}^{2}}}\phi(\Box+m^{2})e^{-\frac{1}{2}\frac{\Box+m^{2}}{\Lambda_{NL}^{2}}}\phi-\frac{1}{2}\varphi(\phi)\frac{\Box+m^{2}}{I-e^{\frac{\Box+m^{2}}{\Lambda_{NL}^{2}}}}\varphi(\phi)-V(\phi+\varphi(\phi))
\end{equation}
and let us neglect the mass parameter, assumed $m<<\Lambda_{NL}$ (this assumption will not be important for our arguments):
\begin{equation}
\label{L3}
\mathcal{L}=-\frac{1}{2}e^{-\frac{1}{2}\frac{\Box}{\Lambda_{NL}^{2}}}\phi(\Box)e^{-\frac{1}{2}\frac{\Box}{\Lambda_{NL}^{2}}}\phi-\frac{1}{2}\varphi(\phi)\frac{\Box}{I-e^{\frac{\Box}{\Lambda_{NL}^{2}}}}\varphi(\phi)-V(\phi+\varphi(\phi))
\end{equation}
where $\varphi(\phi)$ is a functional containing the derivatives of $\phi$ rather than polynomial terms. 
The associated generating function can be formally written as  
\begin{equation}
\label{PathInt}
Z[J]=\int \mathcal{D}\phi \,m[\phi]exp\left(-\int d^{4}x(\mathcal{L}-J\phi) \right)
\end{equation}
where $J$ is the current, $m[\phi]$ the path integral measure. 
 We expand around the classical solution $\phi_{0}$,
 $\phi=\phi_{0}+\phi_{1}$.
 Integrating by part, we can rewrite inside the integral
 $$\mathcal{L}(\phi)\rightarrow \mathcal{L}(\phi_{0})-\phi_{1}M\delta(x-y)+\frac{1}{2}\phi_{1}H\phi_{1}+\mathcal{L}_{int}(\phi_{1})$$
where $M\delta(x-y)$ is the source in the Equation of Motion for $\phi$, $H$ is an infinite derivative operator.
Another useful operation inside the integral is $$J(x)\phi(x)\rightarrow (J(x)-M\delta(x-y))\phi_{1}(x)$$
As a consequence, the path integral can be rewritten as
\begin{equation}
\label{Path2}
Z[J]\sim \int dRd^{4}y \mathcal{Y}(R,y)\frac{e^{-S[\phi_{0}(R,y)]}}{\sqrt{Det\left( H_{R,y}\right)}}e^{-S_{int}\left[\frac{\delta}{\delta J}\right]}
e^{\frac{1}{2}\int d^{4}x_{1}d^{4}x_{2}J(x_{1})\Delta_{R,y}(x_{1},x_{2})J(x_{2})}
\end{equation}
where $R,y$ are the coordinates parametrizing the classicalon solution, 
$\mathcal{Y}(R,y)$ is a measure correspondent to classicalon and the starting $m[\phi]$, 
$\Delta_{R,Y}$ is the inverse function of $H_{R,y}$.
Let us note that $R$ parametrizes the radius of the classicalon. 

All relevant informations about the classicalon are contained in 
the "Green problem" of the propagator $\Delta_{R,y}$:
\begin{equation}
\label{prop2}
H_{R,y}\Delta_{R,y}(\zeta,z)=\delta(\zeta-z)
\end{equation}

Despite to the technical difficulties to find out an
 explicit expression for an infinite derivative operator, 
 it is enough for our purposes to study the asymptotic limits 
 of $H_{R,y}$. 
 In particular, one can argue that for $r\rightarrow \infty$, 
we are OUT of interactions so that the effective lagrangian is asymptotically 
\begin{equation}
\label{L4}
\mathcal{L}\rightarrow -\frac{1}{2}e^{-\frac{1}{2}\frac{\Box}{\Lambda_{NL}^{2}}}\phi \, \Box \, e^{-\frac{1}{2}\frac{\Box}{\Lambda_{NL}^{2}}}\phi
\end{equation} 
 $$= -\left(1-\frac{\Box}{2\Lambda_{NL}^{2}}+\frac{\Box^{2}}{4\Lambda_{NL}^{4}}+...+(-1)^{n}\frac{\Box^{n}}{2^{n}\Lambda_{NL}^{2n}}\right)\, \phi\, \Box\, \left(1-\frac{\Box}{2\Lambda_{NL}^{2}}+\frac{\Box^{2}}{4\Lambda_{NL}^{4}}+...+(-1)^{n}\frac{\Box^{n}}{2^{n}\Lambda_{NL}^{2n}}\right) \phi 
$$
The corresponding Eulero-Lagrange equation can be rewritten as an expansion series 
\begin{equation}
\label{EoM}
\sum_{m=0}^{\infty}A_{m}\Box^{m+1} \phi=0
\end{equation}
where the precise form of $A_{m}$ is not reported because of
not important for our following arguments $(A_{0}=1)$.
 Clearly, in the "deep IR" limit the equation becomes just $\Box \phi=0$. 
 In fact one can perform a Fourier transform of (\ref{EoM}) so that
 \begin{equation}
\label{EoM}
lim_{p\rightarrow 0}\sum A_{m}p^{2m+2} \phi \sim p^{2}\phi+O(p^{4})\phi
\end{equation}
This can be also easily understood as a limit $\Box \rightarrow 0$ of (\ref{L4}): $e^{-\Box/2\Lambda_{NL}^{2}}\rightarrow 1$, $\mathcal{L}\rightarrow -\frac{1}{2}\phi \Box \phi$. 

As a consequence, for $r\rightarrow \infty$,
$H_{R,y}$ can be rewritten in spherical coordinates as
\begin{equation}
\label{rinfty}
H_{R,y}\rightarrow -\sum_{k=0}^{2n+1}\frac{(2n+1)!}{k!(2n+1-k)!} \frac{\partial^{4n+2-2k}}{\partial r^{4n+2-2k}}\sum_{j=0}^{k}\frac{(-1)^{k-j}3^{j}k!}{j!(k-j)!}\frac{1}{r^{j}}\frac{\partial^{j}}{\partial r^{j}}\frac{L^{2(k-j)}}{r^{2(k-j)}}
\end{equation} 
$$-\sum_{k=0}^{2n+1}\frac{(2n+1)!}{k!(2n+1-k)!} \frac{\partial^{4n+2-2k}}{\partial r^{4n+2-2k}}\sum_{j=0}^{k}\frac{(-1)^{k-j}3^{k-j}k!}{j!(k-j)!}\frac{L^{2j}}{r^{2j}}\frac{1}{r^{k-j}}\frac{\partial^{k-j}}{\partial r^{k-j}}$$
$$-\sum_{k=0}^{2n+1}\frac{(2n+1)!}{k!(2n+1-k)!} \frac{L^{4n+2-2k}}{r^{4n+2-2k}}\sum_{j=0}^{k}\frac{(-1)^{k-j}3^{k-j}k!}{j!(k-j)!}\frac{\partial^{2j}}{\partial r^{2j}}\frac{1}{r^{k-j}}\frac{\partial^{k-j}}{\partial r^{k-j}}$$
$$-\sum_{k=0}^{2n+1}\frac{(2n+1)!}{k!(2n+1-k)!} \frac{L^{4n+2-2k}}{r^{4n+2-2k}}\sum_{j=0}^{k}\frac{(-1)^{k-j}3^{j}k!}{j!(k-j)!}\frac{1}{r^{j}}\frac{\partial^{j}}{\partial r^{j}}\frac{\partial^{2(k-j)}}{\partial r^{2(k-j)}}$$
$$-\sum_{k=0}^{2n+1}\frac{3^{2n+1}(2n+1)!}{k!(2n+1-k)!} \frac{1}{r^{2n+1-k}}\frac{\partial^{2n+1-k}}{\partial r^{2n+1-k}}\sum_{j=0}^{k}\frac{(-1)^{k-j}k!}{j!(k-j)!}\frac{\partial^{2j}}{\partial r^{2j}}\frac{L^{2(k-j)}}{r^{2(k-j)}}$$
$$-\sum_{k=0}^{2n+1}\frac{3^{2n+1}(2n+1)!}{k!(2n+1-k)!} \frac{1}{r^{2n+1-k}}\frac{\partial^{2n+1-k}}{\partial r^{2n+1-k}}\sum_{j=0}^{k}\frac{(-1)^{k-j}k!}{j!(k-j)!}\frac{L^{2j}}{r^{2j}}\frac{\partial^{2(k-j)}}{\partial r^{2(k-j)}}$$
where $L^{2}=-\frac{1}{2}(x_{\mu}\partial_{\nu}-x_{\nu}\partial_{\mu})^{2}$.
Clearly, also in the asymptotic limit, it seems
technically difficult to find out a resolvent operator $\Delta_{R,y}$.
On the other hand, in the deep IR limit, the expression $(\ref{rinfty})$ 
has a leading term that is nothing but a derivative of N-th order, 
where $N=4n+2$. This implies that in the momentum space, 
$\Delta_{R,y}$ will trivially reduce to 
\begin{equation}
\label{reduce}
lim_{N\rightarrow \infty, p\rightarrow \infty}\Delta_{R,y}\rightarrow p^{-2}+O(p^{-N})
\end{equation}
$(N>3)$, or 
\begin{equation}
\label{reducea}
lim_{p\rightarrow \infty}\Delta_{R,y}\rightarrow p^{-2}\sum_{N=0}^{\infty}b_{N}p^{-N}
\end{equation}
(where $b_{N}$ is an understood convolution of coefficients in (\ref{rinfty}) not important for our purposes).
In other words, the classical configuration $\phi_{0}$ 
has to satisfy the asymptotic Equation of motion 
\begin{equation}
\label{EoM}
\sum_{m=0}^{\infty}A_{m}\Box^{m+1}\phi_{0}=M\delta(\zeta-z)
\end{equation}

On the other hand, one can consider the UV asymptotic limit of our model, $r\rightarrow 0$. 

In UV regime, $\Box \rightarrow \infty$ so that the relevant lagrangian becomes  
 \begin{equation}
\label{L5}
\mathcal{L}\rightarrow +\frac{1}{2}\varphi(\phi)\frac{\Box}{e^{\frac{\Box}{\Lambda_{NL}^{2}}}}\varphi(\phi)-V(\phi+\varphi(\phi))
\end{equation} 
 where the relevant functional $\varphi$ is $\varphi(\phi)\rightarrow 3\lambda \frac{e^{\frac{\Box}{\Lambda_{NL}^{2}}}}{\Box}\phi^{3} ...$.
 This lagrangian can be explicitly rewritten as an expression of the following leading terms:
 \begin{equation}
 \label{L6}
 \mathcal{L}\rightarrow  +\frac{9\lambda^{2}}{2}
 \frac{1}{n!\Lambda_{NL}^{2n}}\Box^{n-1}\phi^{6}-\frac{81\lambda^{4}}{2}\left(    \frac{1}{n!\Lambda_{NL}^{2n}}\Box^{n-1} \phi^{3}\right)^{4}  
 \end{equation}
 $$-27\lambda^{4}\phi \left(\frac{1}{n!\Lambda_{NL}^{2n}}\Box^{n-1} \phi^{3}\right)^{3}-\frac{54\lambda^{2}}{4}\phi^{2} \left(\frac{1}{n!\Lambda_{NL}^{2n}}\Box^{n-1} \phi^{3}\right)^{2}-\lambda \phi^{3} \left(\frac{1}{n!\Lambda_{NL}^{2n}}\Box^{n-1} \phi^{3}\right)+...$$
 
 As a consequence the asymptotic limit of $H_{R,y}$ for $r\rightarrow 0$, in spherical coordinates, has a form 
 \begin{equation}
 \label{Hzero}
 H_{R,y}\rightarrow -\sum_{k}^{N}\alpha_{Nk}\frac{\partial^{2k}}{\partial r^{2k}}\sum_{j}^{k}\beta_{kj}\frac{3^{j}}{r^{j}}\frac{\partial^{j}}{\partial r^{j}}\frac{L^{2(k-j)}}{r^{2(k-j)}}
 \end{equation}
 $$-\sum_{k}^{N}\alpha_{Nk}\frac{\partial^{2k}}{\partial r^{2k}}\sum_{j}^{k}\beta_{kj}\frac{L^{2j}}{r^{2j}}\frac{3^{k-j}}{r^{k-j}}\frac{\partial^{k-j}}{\partial r^{k-j}}$$
 $$-\sum_{k}^{N}\alpha_{Nk} \frac{L^{2k}}{r^{2k}}\sum_{j}^{k}\beta_{kj} \frac{\partial^{2j}}{\partial r^{2j}}\frac{3^{k-j}}{r^{k-j}}\frac{\partial^{k-j}}{\partial r^{k-j}}$$
 $$-\sum_{k}^{N}\alpha_{Nk} \frac{L^{2k}}{r^{2k}}\sum_{j}^{k}\beta_{kj} \frac{3^{j}}{r^{j}}\frac{\partial^{j}}{\partial r^{j}}\frac{\partial^{2(k-j)}}{\partial r^{2(k-j)}}$$
 $$-\sum_{k}^{N}\alpha_{Nk}\frac{3^{k}}{r^{k}}\frac{\partial^{k}}{\partial r^{k}}\sum_{j}^{k}\beta_{kj}\frac{L^{2j}}{r^{2j}}\frac{\partial^{2(k-j)}}{\partial r^{2(k-j)}}$$
 $$ -\sum_{k}^{N}\alpha_{Nk}\frac{3^{k}}{r^{k}}\frac{\partial^{k}}{\partial r^{k}}\sum_{j}^{k}\beta_{kj}\frac{\partial^{2j}}{\partial r^{2j}}\frac{L^{2(k-j)}}{r^{2(k-j)}}$$
 
 where $\alpha_{Nk},\beta_{kj}$ are combinations of factorials, not explicitly reported: they will be not important for our purposes.
 As a consequence, in deep UV regime, $H_{R,y}$ is a combination of all $\partial^{M}/\partial r^{M}$ and $1/r^{P}$
 satisfying the constraint $M+P=2N$:
 \begin{equation}
 \label{MPN}
lim_{N\rightarrow \infty, r\rightarrow 0} H_{R,y} \sim -lim_{N\rightarrow \infty}\sum_{M,P;M+P=2N}B_{MP}\frac{1}{r^{P}}\frac{\partial^{M}}{\partial r^{M}}
 \end{equation}
 where $B_{MP}$ is a combination of $\alpha,\beta$ coefficients. 
From this, we can see how divergent is the operator $H_{R,y}$ in the deep UV regime. 
 As a consequence, we can find a trivial solution $\Delta_{R,y}$ for the problem Eq.(\ref{prop2}) in the asymptotic limit $r\rightarrow 0$:
just the trivial distribution $\Delta_{R,y} \rightarrow 0$. 
 
Now, we can reconstruct a  $\Delta_{R,y}$ 
connecting the two asymptotic solutions $r\rightarrow \infty$ 
and $r\rightarrow 0$ discussed above.
The resolvent $\Delta_{R,y}$, in the momentum space, will have a form 
\begin{equation}
\label{FinalResult}
lim_{N\rightarrow \infty}\Delta_{R,y} \sim f(p)\mathcal{F}(p_{*}-p)
\end{equation}
where
$f(p)$ is a monotonically decreasing function with a limit $p\rightarrow \infty$ converging to a power series 
$p^{-2}\sum_{N=0}b_{N}p^{-N}$ of (\ref{reducea}).
$\mathcal{F}(p_{*}-p)$ is a distribution that has to satisfy the following asymptotic proprieties: 

for $p<hp_{*}$  
$$f(p)\mathcal{F}(p_{*}-p)=0$$
for $p>kp_{*}$ 
$$f(p)\mathcal{F}(p_{*}-p)=qf(p)$$
where $q>0$ is a constant, 
with $k,h$ two real numbers satisfying $k\geq h>0$.
In the following discussion, we will assume $q=1$:
simply we can redefine $f(p)$ so that $q$ is just embedded 
in this one. Let us note that, at priori, $f(p)$ can be a function defined in the range $(-\infty,p^{*}]$,
{\it i.e} not necessary in $(-\infty,\infty)$ or $(-\infty,0]$. 
These are all the general informations that we can get by our "deep" asymptotic limits.
Among the possible solutions, one can propose as an {\it ansatz}
that 
$$\mathcal{F}(p_{*}-p)=\Theta(p^{*}-p)$$
where $\Theta$ is the Heaviside distribution. 
In fact, this case satisfies all conditions mentioned above. 
In particular, if $k=h=1$ one obtain this result. 
Let us define $R=1/p_{*}$.
This definition is useful to get the regime in which surely 
one can consider a large class of $\mathcal{F}(p_{*}-p)$ 
as a Heaviside distribution $\Theta(p_{*}-p)$. 
In fact, $R/h<r<R/k$ is the region in which  
a generic $\mathcal{F}$ deviates from an Heaviside step distribution:
we can imagine it as a smoothed Heaviside (around the step).
On the other hand, for $r<<R/h$ and $r>>R/k$, 
$\mathcal{F}$ converges to a Heaviside distribution.

The fact that we have chosen the same labels 
 $R$ for the classicalon radius
as well as for $R=1/p_{*}$ is not a coincidence:
the physical interpretation of $R$ inside the distribution 
$\mathcal{F}$ is exactly the classicalon radius.  
In fact, the transition of $\mathcal{F}$ is related to the presence 
of a classical configuration of radius $R$. 
At this point, one can argue that the particular shape of $\mathcal{F}$
around the Heaviside distribution has not a particular physical importance
for our purposes: we are only interested to get qualitative 
proprieties of classicalization in scattering amplitudes.
In a broad analogy,
this is like the case of shape functions for atomic nuclei 
in which one can define an average radius $R$, with a monotonically decreasing  
behavior after $R$.

Let us discuss the important implications of these formal results
on scattering amplitudes. 
 In particular, we are interested to a $2\rightarrow 2 $ scattering as a
significant example. So, let us consider the 4-correlator (4-Green's function):
\begin{equation}
\label{corr}
\mathcal{G}(z_{1},z_{2},z_{3},z_{4})=\frac{1}{Z}\int dRd^{4}y \mathcal{Y}(R,y)\frac{e^{-S[\phi_{0}(R,y)]}}{\sqrt{Det\left( H_{R,y}\right)}}e^{-S_{int}}
\end{equation}
$$\times \left[\partial_{\mu}\Delta_{R,y}(z_{1},z)\partial^{\mu}\Delta_{R,y}(z_{2},z)\partial_{\nu}\Delta_{R,y}(z_{3},z)\partial^{\nu}\Delta_{R,y}(z_{4},z)+permutations\right]$$
where $Z$ is the partition function (\ref{Path2})
 At this point, we can find out some important qualitative proprieties 
 of $\mathcal{G}(z_{1},z_{2},z_{3},z_{4})$
 contained in our solution $\Delta_{R,y}$ of the form (\ref{FinalResult}). 
 Inserting the Heaviside-like distributions into (\ref{corr}), we will emerge a cutoff $R$ inside the Green correlator. 
So that, from these very general and 
simple considerations, we can conclude that 
the maximal momentum probed by the $2\rightarrow 2$ scattering can be just $p_{*}=1/R$.
As a consequence the 4-amplitude is 
\begin{equation}
\label{fouramp}
lim_{p\rightarrow \infty, n\rightarrow \infty}\mathcal{M}_{\phi\phi\rightarrow \phi\phi}\leq a_{n}\frac{p_{*}^{2n}}{\Lambda_{NL}^{2n}}
\end{equation}
We know the final result of our amplitude at 1-loop 
(\ref{Amplitude1}). Substituting $s=p_{*}^{2}\leq \Lambda_{NL}^{2}$ in the series, 
the convergence can be manifestly shown. Infact, $a_{n}< 1/(n+1)!n$, 
with $a_{n}$ the coefficient of our series: 
for the direct comparison test, 
convergence of $\sum_{n}1/(n+1)!n$ implies convergence of $\sum_{n}a_{n}$.
For example,  in s-channel, we have
\begin{equation} 
\label{rewa2a}
lim_{s\rightarrow \Lambda^{2}}\mathcal{A}\rightarrow \frac{9\lambda^{2}}{4\pi^{2}}\sum_{n=0}^{\infty}\frac{1-2^{-n}}{(n+1)!n}
(\tilde{s}^{*})^{n}
\end{equation}
and the following bound:
\begin{equation} 
\label{rewa2a}
lim_{s> \Lambda^{2}}\mathcal{A}\leq \frac{9\lambda^{2}}{4\pi^{2}}\sum_{n=0}^{\infty}\frac{1-2^{-n}}{(n+1)!n}
(\tilde{s}^{*})^{n}
\end{equation}
where $\tilde{s}$ is the adimensional Mandelstam variable 
$\tilde{s}=s/\Lambda_{NL}^{2}$, 
so that 
$\tilde{s}^{*}=s/\Lambda_{NL}^{2}=(p^{*})^{2}/\Lambda_{NL}^{2}$
is the cut-off value in the s-channel. 
The convergence of this amplitude to a finite one is manifest for $\tilde{s}^{*}\leq 1$. 
Analogous considerations are valid for t- and u- channels. 
More surprisingly, the amplitude (\ref{Amplitude1}) can converge also for 
$\tilde{s}^{*}=const>1$, thanks to the factor $1/(n+1)!$.

\subsection{Further comments and implications}

In this section, we would like to briefly comment 
about some implications of classicalization
in non-local models. These aspects 
will deserve deeper analysis and future investigations beyond the purposes of this paper. 

1) Our result can be extended for vector-bosons' scatterings. 
As shown in \cite{W1,W2,Moffat,W3,Evens}, a non-local gauge theory can be formulated.
Scattering like $VV\rightarrow VV$, where $V$ is a vector boson, 
for $E>>\Lambda_{NL}$ behaves similarly to
the one considered here: an infinite series of divergences 
will emerge \cite{Joglekar1,Joglekar2,Addazi:2015dxa}.

2) UV fate of non-local models, {\it i.e} 
how to decide if a non-local theory is Wilson-like UV completed
or Dvali-like one. We can argue that this problem 
is not different in our case with respect to local QFTs. 
These aspects were just discussed in various papers cited above.
Because of this, 
we will not discuss this issue in our paper.

3) Classicalization can help non-local models to 
eliminate acausal divergences in scatterings,
{\it i.e} it is an unitarization and causalization of scatterings.
However, divergences coming from radiative corrections 
of propagators or vacuum polarization diagrams seem 
to remain still alive! These kinds of divergences 
were considered in our paper \cite{Addazi:2015dxa}
(in contest of $\mathcal{N}=1$ susy non-local models). 
These divergences can have observable effects 
in the running coupling constants. Clearly, 
these are suppressed as a powers $\Lambda_{NL}^{-n}$
so that these effects are negligible for $E<<\Lambda_{NL}$. 
 $\mathcal{N}=1$ (rigid) supersymmetry seems to help to cancel
an infinite number of radiative divergences, as explicitly shown in \cite{Addazi:2015dxa},
even if not all the infinite ones! (Curiously) $\mathcal{N}>1$ susy non-local models
were not studied in literature. One can conjecture that for $\mathcal{N}=2$
susy non-local models, more divergences can be eliminated. 

4) EWMKEJ model is manifestly Lorentz invariant and CPT invariant 
at three level. However, one can argue that classicalization can
be a more general procedure for a more general class of non-local models
without Lorentz and CPT invariance. This can be strongly motivated by 
QFT in non-commutative geometries in which non-local 
terms in matter generically emerge \footnote{For a recent new model of non-commutative QFT see \cite{Kawamoto:2015qla}. }
\footnote{An intriguing area to explore could be non-local models in which Lorentz Invariance
is an emergent symmetry. See \cite{SLIV1,SLIV2,SLIV3} for examples of quantum field theories in which Lorentz invariance
is not a fundamental symmetry.}.

5) In our considerations, we have not considered gravity. 
However, one can argue that these results can be also extended 
for non-local modifications of General-Relativity \cite{NONLOCALGRAVITY1,NONLOCALGRAVITY2,
NONLOCALGRAVITY3,NONLOCALGRAVITY4,NONLOCALGRAVITY5,NONLOCALGRAVITY6,NONLOCALGRAVITY7,
NONLOCALGRAVITY8,NONLOCALGRAVITY9,
NONLOCALGRAVITY10,NONLOCALGRAVITY11,NONLOCALGRAVITY12,NONLOCALGRAVITY13,NONLOCALGRAVITY14,NONLOCALGRAVITY15}\footnote{We would like to mention that recently a study of external geodetic stability 
in particular branches of massive gravity \cite{Addazi:2014mga}.
 In subregions of parameters of these models, naked singularities can 
exist. This can be connected to the existence of new items called {\it frizzyballs} 
\cite{Addazi:2015gna,Addazi:2015hpa}.}.
In these model, GR singularities are removed, but 
quantization is still inconsistent. However, 
classicalons' formation in graviton-graviton scatterings
can unitarize 
a non-local extension of GR. 
As notice by Dvali and collaborators, a classicalon 
can be nothing but a black hole in gravitational scatterings \cite{DvaliBH}. A Black hole can be formed 
in graviton-graviton scatterings, unitarizing a gravitational theory. 

6) In the case of $\Lambda_{NL}\simeq 100\, \rm TeV$, one can speculate 
about implications for UHECR or for future colliders. 
In particular: i) classicalon resonance can give characteristic signatures 
in collisions, as proposed in papers about classicalizations cited above; ii) as mentioned above, polynomial corrections 
to the cross sections can be a clear signal beyond a local quantum field theory
\cite{Joglekar1,Joglekar2, Joglekar3, Joglekar4, Mazumdar1}.

\section{Conclusion and remarks}

In this paper, 
we have shown how classicalization
can cure acausal divergences 
of non-local QFT coming in "trans-non-local" 
limit. In particular, we have discussed a particular 
class of non-local QFT for a scalar field
well studied in litterature cited above. 
We have explicitly shown how the formation of a classicalon 
can avoid acausalities in scatterings for an energy higher
than the non-locality scale.  
We have also discussed possible implications
for gauge theories, gravity, cosmology, UHECR and future $100\, \rm TeV$-colliders.
We conclude that classicalization seems a natural UV completion 
of non-local quantum field theories, unitarizing and causalizing their scattering amplitudes.

\vspace{1cm} 

{\large \bf Acknowledgments} 
\vspace{3mm}

I would like to thank Giampiero Esposito, Leonardo Modesto, Anumap  Mazumdar 
and Silvia Vada
for discussions on these aspects. 
My work was supported in part by the MIUR research
grant "Theoretical Astroparticle Physics" PRIN 2012CPPYP7.

% *** Biliography ***

\end{document}